\documentstyle[aps%,twocolumn%
,epsfig%
,preprint%
]{revtex}
\begin{document}
\draft
\preprint{Guchi-TP-004}
\date{\today%August 1999
}
\title{
Rotating Boson Stars with Large Self-interaction in (2+1) dimensions
}
\author{Kenji~Sakamoto${}^{1,2}$%
\thanks{e-mail: {\tt
b1795@sty.sv.cc.yamaguchi-u.ac.jp,
saka@ccthmail.kek.jp}
% \hfill\break
} and
Kiyoshi~Shiraishi${}^{1,3}$%
\thanks{e-mail: {\tt
shiraish@sci.yamaguchi-u.ac.jp}
% \hfill\break
}}
\address{${}^1$Graduate School of Science and Engineering,
Yamaguchi University\\
Yoshida, Yamaguchi-shi, Yamaguchi 753-8512, Japan
}
\address{${}^2$Institute of Particle and Nuclear Studies,
High Energy Accelerator Research Organization (KEK)\\
Tsukuba, Ibaraki 305-0801, Japan
}
\address{${}^3$Faculty of Science, Yamaguchi University\\
Yoshida, Yamaguchi-shi, Yamaguchi 753-8512, Japan
}
\maketitle

\begin{abstract}
%%%%%%%%%%%%%%%%%%%%%%%%%%%%%%%%%%%%%%%%%%%%%%%%%%%%%%%%%%%%%%%%%%%%%%
Solutions for rotating boson stars in $(2+1)$ dimensional
gravity with a negative cosmological constant are obtained
numerically.
The mass, particle number, and radius of
the $(2+1)$ dimensional rotating boson star are shown.
Consequently we find the region where the stable boson star can exist.
%%%%%%%%%%%%%%%%%%%%%%%%%%%%%%%%%%%%%%%%%%%%%%%%%%%%%%%%%%%%%%%%%%%%%%
\end{abstract}

%\vspace{7mm}
%\vfill
%\eject

%%%%%%%%%%%%%%%%%%%%%%%%%%%%%%%%%%%%%%%%%%%%%%%%%%%%%%%%%%%%%%%%%%%%%%

%%%%%%%%%%%%%%%%%%%%%%%%%%%%%%%%%%%%%%%%%%%%%%%%%%%%%%%%%%%%%%%%%%%%%%
%%%%%%%%%%%%%%%%%%%%%%%%%%%%%%%%%%%%%%%%%%%%%%%%%%%%%%%%%%%%%%%%%%%%%%

Self-gravitating systems have been investigated in various situations.
Boson stars (\cite{BS}\cite{LM} for reviews) have a very simple
constituent,  a complex scalar field which is bound by gravitational
attraction.
Thus the boson star provides us with the simplest model of
relativistic stars.

The solutions for relativistic boson stars are only numerically
obtained in four dimensions. In $(2+1)$ dimensions, static equilibrium
configurations have been argued \cite{CZ} in Einstein gravity with a
negative cosmological constant.
In the previous paper~\cite{rf:3}, we obtained an exact solution for nonrotating boson star
in $(2+1)$ dimensional gravity with a negative cosmological constant.
We consider that the scalar field has a strong
self-interaction.
An infinitely large self-interaction term in the model leads to much
simplications as in the $(3+1)$ dimensional case~\cite{CSW}.

In the present paper, we obtain numerical solutions for a rotating boson
star in $(2+1)$ dimensional gravity with a negative cosmological
constant. We assume that the scalar field has a strong
self-interaction as in the previous paper~\cite{rf:3}.
The rotating boson star in the $(3+1)$ dimensional case has been studied
numerically by F.~D.~Ryan~\cite{Ryan}.
We wish to study the similarity and the difference, between
 the $(2+1)$ dimensional model and the model in other dimensions.
The study of the rotating boson star will lead to a new aspect of gravitating
systems and clarify the similarity and/or the difference among the
other dimensional cases.

%%%%%%%%%%%%%%%%%%%%%%%%%%%%%%%%%%%%%%%%%%%%%%%%%%%%%%%%%%%%%%%%%%%%%%

We consider a complex scalar field with mass $m$ and a quartic
self-coupling constant $\lambda$.
The action for the scalar field coupled to gravity can be
written down as
\begin{equation}
S=\int d^3x\sqrt{-g}\left[
\frac{1}{16\pi G}\left(R+2C\right)-
\left|\nabla_{\mu}\varphi\right|^2-
m^2\left|\varphi\right|^2-
\frac{\lambda}{2}\left|\varphi\right|^4\right],
\label{eq:action}
\end{equation}
where $R$ is the scalar curvature and the positive constant
$C$ stands for the (negative) cosmological constant.
$G$ is the Newton constant.

Varying the action (\ref{eq:action}) with respect to
the scalar field and the metric yields equations of
motion. The equation of motion for the scalar field is
\begin{equation}
\nabla^2\varphi-m^2\varphi-\lambda\left|\varphi\right|^2\varphi=0,
\label{eq:em}
\end{equation}
while the Einstein equation is
\begin{equation}
R_{\mu\nu}-\frac{1}{2}g_{\mu\nu}R=
8\pi G\left[
2{\rm Re}\left(\nabla_{\mu}\varphi^{*}\nabla_{\nu}\varphi-
\frac{1}{2}\left|\nabla_{}\varphi\right|^2 g_{\mu\nu}\right)-
m^2\left|\varphi\right|^2g_{\mu\nu}-\frac{\lambda}{2}
\left|\varphi\right|^4g_{\mu\nu}\right]+Cg_{\mu\nu}.
\label{eq:Ee}
\end{equation}

We assume the three-dimensional metric
for a circularly symmetric spacetime as
\begin{equation}
ds^2=-e^{-2\delta(r)}\Delta(r)dt^2+\frac{1}{\Delta(r)}dr^2+
r^2(d \theta-\Omega(r) dt)^2,
\label{eq:metr}
\end{equation}
where $\delta$, $\Delta$, and $\Omega$ are functions of the radial
coordinate $r$ only. We will later use $\Delta=Cr^2-8GM(r)$, where $M$ is
a function of $r$.

We also assume that the complex scalar has the following dependence on
the coordinates,
\begin{equation}
\varphi=e^{-i\omega t+i\ell \theta}\varphi(r),
\end{equation}
where $\varphi(r)$ is a function of the radial coordinate $r$ only
and $\omega$ and $\ell$ are constants.

Then the equations~(\ref{eq:em}) and (\ref{eq:Ee}) are written as
\begin{eqnarray}
\frac{1}{2r} \frac{d\Delta}{dr}+
\frac{1}{4} r^2 e^{2\delta}
\left(\frac{d\Omega}{dr}\right)^2&=&
-8\pi G\biggl[\Delta \left(\frac{d\varphi}{dr}\right)^2
+\frac{\ell^2}{r^2}\varphi^2
+\frac{e^{2\delta}}{\Delta}
 \left( \omega-\ell\Omega\right)^2 \varphi^2 \nonumber \\
& & \hspace{3cm} +m^2 \varphi^2
+\frac{\lambda}{2} \varphi^4 \biggl]+C, \hspace{1cm} \\
\frac{\Delta}{r} \frac{d\delta}{dr}&=&
-16\pi G\left[\Delta \left(\frac{d\varphi}{dr}\right)^2
+\frac{e^{2\delta}}{\Delta} \left( \omega-\ell\Omega\right)^2
 \varphi^2 \right], \hspace{1cm} \\
-\frac{1}{2r^2}\sqrt{\Delta}
\frac{d}{dr}\left( r^3 e^{\delta}
\frac{d\Omega}{dr}\right)&=&
16\pi G\frac{e^{\delta}}{\sqrt{\Delta} r}\ell
 \left( \omega-\ell\Omega\right)^2 \varphi^2 , \\
\frac{1}{r e^{-\delta}}
\frac{d}{dr}\left( r e^{-\delta} \Delta
\frac{d\varphi}{dr} \right)&=&
\frac{\ell^2}{r^2 } \varphi
-\frac{e^{2\delta}}{\Delta}
 \left( \omega-\ell\Omega\right)^2 \varphi
+m^2\varphi+\lambda \varphi^3.
\end{eqnarray}

For convenience, we rescale the variables as
\begin{equation}
\tilde{r}=mr,~~~\tilde{\varphi}=\sqrt{8\pi G}\varphi,~~~
\Lambda=\frac{\lambda}{8\pi Gm^2},~~~
\tilde{\Omega}=\frac{1}{m} (\Omega- \frac{\omega}{\ell}).
\end{equation}
We also denote $\Delta=Cr^2-8GM=(C/m^2)\tilde{r}^2-8GM$.

Using these variables, we can rewrite the equations as
\begin{eqnarray}
-\frac{1}{2\tilde{r}} \frac{d(8GM)}{d\tilde{r}}+
\frac{1}{4} \tilde{r}^2 e^{2\delta}
\left(\frac{d\tilde{\Omega}}{d\tilde{r}}\right)^2&=&
-\biggl[\Delta \left(\frac{d\tilde{\varphi}}{d\tilde{r}}\right)^2
+\frac{\ell^2}{\tilde{r}^2}\tilde{\varphi}^2
+\frac{e^{2\delta}}{\Delta}
 \ell^2 \tilde{\Omega}^2 \tilde{\varphi}^2
+\tilde{\varphi}^2
+\frac{\Lambda}{2} \tilde{\varphi}^4
\biggl], \hspace{1cm} \\
\frac{\Delta}{\tilde{r}} \frac{d\delta}{d\tilde{r}}&=&
-2\left[\Delta \left(\frac{d\tilde{\varphi}}{d\tilde{r}}\right)^2
+\frac{e^{2\delta}}{\Delta} \ell^2 \tilde{\Omega}^2
 \tilde{\varphi}^2 \right], \hspace{1cm} \\
-\frac{1}{2\tilde{r}^2}
\frac{d}{d\tilde{r}}\left( \tilde{r}^3 e^{\delta}
\frac{d\tilde{\Omega}}{d\tilde{r}}\right)&=&
-2\frac{e^{\delta}}{\Delta \tilde{r}} \ell^2
 \tilde{\Omega}~\tilde{\varphi}^2 , \\
\frac{1}{\tilde{r} e^{-\delta}}
\frac{d}{d\tilde{r}}\left( \tilde{r} e^{-\delta} \Delta
\frac{d\tilde{\varphi}}{d\tilde{r}} \right)&=&
\frac{\ell^2}{\tilde{r}^2 } \tilde{\varphi}
-\frac{e^{2\delta}}{\Delta} \ell^2 \tilde{\Omega}^2 \tilde{\varphi}
+\tilde{\varphi}+\Lambda \tilde{\varphi}^3.
\end{eqnarray}
%%%%%%%%%%%%%%%%%%%%%%%%%%%%%%%%%%%%%%%%%%%%%%%%%%%%%%%%%%%%%%%%%%%%%%

Next we rescale the variables again, to study the limit of
large self-inteaction.
New variables are:
\begin{equation}
r_{*}=\tilde{r}/\sqrt{\Lambda},~~~
\varphi_{*}=\sqrt{\Lambda}\tilde{\varphi},~~~
L^2=\frac{\ell^2}{\Lambda},~~~
\Omega_*=\sqrt{\Lambda}\tilde{\Omega},
\end{equation}
and $\Delta=Cr^2-8GM=C_*r^2_*-8GM$, where $C_*=C\Lambda/m^2$.

Then the equations can be written as
\begin{eqnarray}
-\frac{1}{2r_*} (8GM')
 +\frac{1}{4} r_*^2 e^{2\delta}(\Omega_*')^2&=&
\biggl[\frac{1}{\Lambda} \Delta (\varphi_*')^2
+\frac{L^2}{r_*^2} \varphi_*^2
+\frac{e^{2\delta}}{\Delta}L^2\Omega^2_* \varphi_*^2
+\varphi_*^2+ \frac{1}{2} \varphi_*^4\biggl],\hspace{1cm}\\
\frac{\Delta}{r_*}\delta'&=&
-2\left[\frac{1}{\Lambda} \Delta (\varphi_*')^2
+\frac{e^{2\delta}}{\Delta}L^2\Omega^2_* \varphi_*^2\right],\hspace{1cm}\\
-\frac{1}{2r_*^2} \left( r_*^3 e^{\delta} \Omega_*' \right)'&=&
2\frac{e^{\delta}}{\Delta r_*}L^2 (-\Omega_*) \varphi_*^2,\\
\frac{1}{\Lambda} \frac{1}{r_* e^{-\delta} }
\left(r_* e^{-\delta} \Delta \varphi_*' \right)'&=&
\frac{L^2}{r_*^2}\varphi_*
-\frac{e^{2\delta}}{\Delta}L^2 \Omega_*^2 \varphi_*
+\varphi_*+\varphi_*^3,\hspace{1cm}
\end{eqnarray}
where ${}'$ denotes the derivative with respect to $r_{*}$.

Now we consider the limit of large self-interaction.
The study of boson stars
with the finite value of self-coupling is very difficult.
Thus we simplify the field equations by the limit of large self-interaction.
For the limit of large self-coupling $\Lambda\rightarrow\infty$,
these equations will be reduced to%
%corrected(1)
\footnote{
For a finite $C_*$, the actual value of $C=m^2C_*/\Lambda$ becomes
infinitely small if the limit $\Lambda=\infty$ is taken literally.
We can however interpret the limit as an approximation of a
{\it large}
self-coupling, and $\Lambda$ is not simply taken as a mathematical
infinity.
}
\begin{eqnarray}
\frac{1}{2r_*} \bar{M}'&=&
\frac{1}{4L^2} r_*^2e^{2\delta} (\bar{\Omega}')^2
+\frac{L^2}{r_*^2} \varphi_*^2
+\frac{e^{2\delta}}{\Delta}\bar{\Omega}^2 \varphi_*^2
+\varphi_*^2+ \frac{1}{2} \varphi_*^4,
\label{eq:rot1} \hspace{1cm} \\
\frac{1}{r_*}\delta'&=&
-2\frac{e^{2\delta}}{\Delta^2}\bar{\Omega}^2 \varphi_*^2,
\label{eq:rot2} \\
\frac{1}{r_*} \left( r_*^3 e^{\delta} \bar{\Omega}' \right)'&=&
4\frac{e^{\delta}}{\Delta}L^2\bar{\Omega} \varphi_*^2,
\label{eq:rot3} \\
\varphi_*^3&=&
\varphi_*\left(\frac{e^{2\delta}}{\Delta}\bar{\Omega}^2
-\frac{L^2}{r_*^2}-1 \right),
\label{eq:rot4}
\end{eqnarray}
where $\bar{M}=8GM$ and $\bar{\Omega}=L\Omega_*$.
%%%%%%%%%%%%%%%%%%%%%%%%%%%%%%%%%%%%%%%%%%%%%%%%%%%%%%%%%%%%%%%%%%%%%%

Now, let us solve the set of an algebraic equation and differential
equations~(\ref{eq:rot1}-\ref{eq:rot4}).

We can solve Eq.(\ref{eq:rot4}) for $\varphi_*$ easily.
A trivial solution is
\begin{equation}
\varphi_*^2=0.
\end{equation}
This corresponds to the region of no bosonic matter field, i.e.,
the vacuum outside a star.
The interior solution for $\varphi_*$ is
\begin{equation}
\varphi_*^2=\frac{e^{2\delta}}{\Delta} \bar{\Omega}^2
-\frac{L^2}{r_*^2}-1.
\end{equation}
This describes the configuration of the boson field inside a star.

The term $L^2/r_*^2$ yields a remarkable consequence.
For nonzero $L^2$, any positive value for
the solution become impossible in the vicinity of the origin.
Thus, for a sufficiently small
value for $r_*$, the solution for $\varphi_*^2$
must be $\varphi_*^2=0$.
In other words, a rotating boson star has a vacuum ``hole''
in its central region!
We note that the rotating boson star in (3+1) dimensions
also has a similar structure in the central region~\cite{Ryan}.

If we require that there is no singularity in the central vacuum region, the spacetime must
be a usual anti-de Sitter spacetime.
Therefore in the central region
\begin{equation}
\Delta=1+Cr^2=1+C_*r_*^2.
\end{equation}
This determines the value of $\bar{M}$ in the central region as $\bar{M}=-1$.

Unfortunately, a set of equations~(\ref{eq:rot1}-\ref{eq:rot4}) cannot be solved
analytically.
Thus, we solve these equations numerically here.

To solve the full solution numerically, we must specify the boundary conditions.
We consider that the central vacuum region is the range $0<r_*<r_{*i}$.
The value for $r_{*i}$ gives the location of the inside boundary of
a rotating boson star. Here
\begin{equation}
\bar{M}(r_{*i})=-1,
\end{equation}
from the previous consideration.
Obviously $\delta$ has a constant value in the vacuum regions.
Thus the value for $\delta$ in the vacuum regions can be additively modified by
redefinition of the time coordinate, etc.. Therefore we take
\begin{equation}
\delta(r_{*i})=0,
\end{equation}
without any loss of generality.
Because of the continuity of the solution for $\varphi_*$,
a condition is required:
\begin{equation}
\frac{\bar{\Omega}^2_i}{C_*r^2_{*i}+1}
-\frac{L^2}{r_{*i}^2}-1=0,
\end{equation}
where $\bar{\Omega}_i=\bar{\Omega}(r_{*i})$.

Now we will solve the field equations numerically from $r_*=r_{*i}$.
In the numerical solutions, $\varphi_*^2$ becomes zero again
at $r_*=r_{*o}>r_{*i}$.
We consider that the region $r_*>r_{*o}$ is outside of a rotating boson star,
that is, the vacuum region where $\varphi_*^2=0$.
In this region, we can analytically solve the equations by using the boundary value
for variables at $r_*=r_{*o}$.
In the outside vacuum region, the metric is described by BTZ solution \cite{BTZ}.

\begin{figure}[htbp]
\centering
\epsfbox{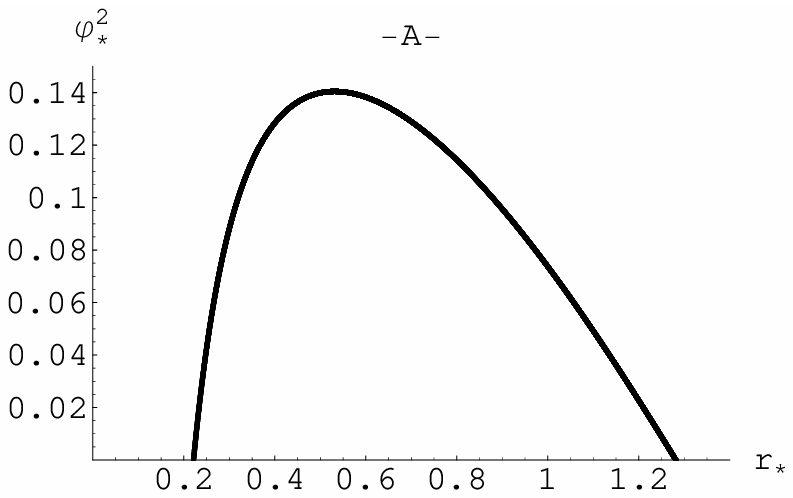}
\epsfbox{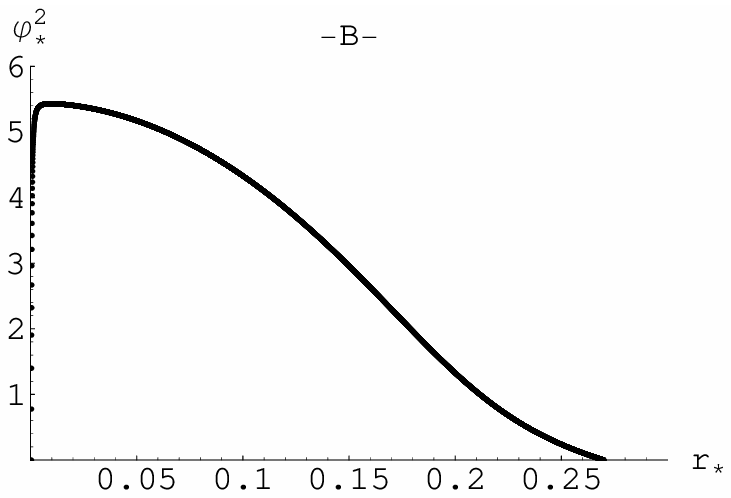}
\epsfbox{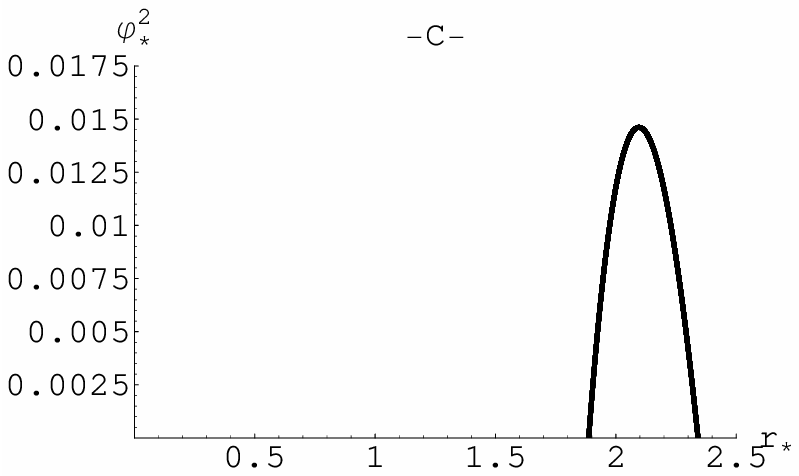}
\caption{$\varphi_*^2$ as a function of
 $r_*$, where (A) $C_*=0.1,~\bar{\Omega}_i=1.1,~L=0.1$,
(B) $C_*=2.526,~\bar{\Omega}_i=2.54,~L=0.001$ and
(C) $C_*=0.1,~\bar{\Omega}_i=1.45,~L=1.4$.}
\label{fig1}
\end{figure}

By these conditions, we can solve the field equations numerically.
Now the arbitary parameter set is three conditions, $\bar{\Omega}_i$, $C_*$ and $L$.
Numerical solutions for some parameter sets are shown Fig. \ref{fig1}.

In Fig.~\ref{fig1} $\varphi_*^2$ as a function of the radial coordinate $r_*$ is shown.
As seen before, we see that the boson star has a vacuum hole in the center of the star.
 $L$ is the parameter that indicates the magnitude of the rotation.
For a small $L$, the configuration of the boson star is similar
to the nonrotating boson star. This nonrotating boson star has already been studied~\cite{rf:3}.
For a large $L$, the boson star has a large vacuum hole and a large radius.
The region where the matter exists becomes narrow.

We cannot, however,  caluculate numerical values of $\varphi_*^2$, $\Delta$,
$\delta$, and $\bar{\Omega}$ for {\it all} set of the conditions $\bar{\Omega}_i$, $L$ and $C_*$.
For some parameter sets, $\varphi_*^2$ diverges before connecting the outside vacuum region.
Scalar curvature of the rotating boson star is the following:
\begin{equation}
R=-16\pi G \left[ 2m^2 \varphi^2+\frac{\lambda}{2} \varphi^4 \right].
\end{equation}
If $\varphi_*^2$ diverge, scalar curvature also diverge.
Then there is a singularity.
Therefore it is worth noting that the parameter
region for the possible solutions is restricted.

For the possible solutions, we can calculate the physical quantities.
Now, we consider the mass of the boson star.
Since the external solutions of boson stars correspond to BTZ solutions,
the mass of the boson star is identified with the BTZ mass.
We define the total mass of the boson star with the mass of the boson star observed at infinity.
For the outside of the boson star, the equation (\ref{eq:rot1}) and (\ref{eq:rot3}) are shown
\begin{eqnarray}
\frac{1}{2r_*} \bar{M}'&=&\frac{1}{4L^2}r^2_*e^{2\delta}
\left(\bar{\Omega}'\right)^2,\label{610} \\
\frac{1}{r_*}\left(r^3_*e^{\delta}\bar{\Omega}'\right)'&=&0. \label{611}
\end{eqnarray}
From the equation (\ref{611}),
\begin{equation}
r^3_*e^{\delta}\bar{\Omega}'=r^3_{*o}e^{\delta_o}\bar{\Omega}'_o=Const.,
\end{equation}
where $r_*=r_{*o}$ is the external boundary of the boson star, and
$\delta_o=\delta(r_{*o})$,
$\bar{\Omega}_o'=\bar{\Omega}'(r_{*o})$.
Then we find from the equation (\ref{610}),
\begin{eqnarray}
\frac{1}{2r_*} \bar{M}'&=&\frac{1}{4L^2}r^2_*e^{2\delta}
\left(\bar{\Omega}'\right)^2 \nonumber \\
&=&\frac{1}{4L^2}r^2_*e^{2\delta}
\left(\frac{r^3_{*o}e^{2\delta_o}}{r^3_*e^{\delta}}
\bar{\Omega}'_o\right)^2, \\
\bar{M}'&=&\frac{1}{2L^2r^3_*}r^6_{*o}e^{2\delta_o}(\bar{\Omega}'_o)^2.
\end{eqnarray}
We integrate this in the external region of the boson star:
\begin{eqnarray}
\int^{\infty}_{r_{o*}}\bar{M}'dr_*
&=&\frac{1}{2L^2}r^6_{*o}e^{2\delta_o}(\bar{\Omega}'_o)^2\int^\infty_{r_{o*}}
\frac{1}{r^3_*}dr_* \nonumber \\
&=&\frac{r^4_{o*}}{4L^2}e^{2\delta_o}(\bar{\Omega}'_o)^2.
\end{eqnarray}
Here, we note that
\begin{equation}
\int^{\infty}_{r_{o*}}\bar{M}'dr_*=\bar{M}(\infty)-\bar{M}(r_{*o}).
\end{equation}

Thus the total mass of the boson star is given,
\begin{equation}
8GM_{BTZ}=\bar{M}(\infty)=\bar{M}_o+\frac{r_{*o}^4}{4L^2} e^{2\delta_o}
(\bar{\Omega}_o')^2,
\end{equation}
where $\bar{M}_o=\bar{M}(r_{*o})$.

The particle number of the boson star is given,
\begin{eqnarray}
N&=&-\int d^2x \sqrt{-g}~i~\left(
\varphi\nabla^t\varphi^{*}
-\varphi^{*}\nabla^t\varphi\right) \nonumber \\
&=&\int d^2x re^{-\delta}~i~\Biggl[\frac{e^{2\delta}}{\Delta}\left(2~i~\omega\varphi^2\right)+\frac{e^{2\delta}}{\Delta}\Omega\left(-2~i~\ell\varphi^2\right)\Biggl] \nonumber \\
&=&2\pi\int dr\frac{re^{\delta}}{\Delta}2\varphi^2\left(\Omega\ell-\omega\right) \nonumber \\
&=&\frac{1}{2Gm}\int dr_*\frac{r_*\varphi^2_*e^{\delta}\bar{\Omega}}{C_*r_*^2-\bar{M}} \nonumber \\
&=& \frac{1}{2Gm} \int dr_*~\frac{r_*  e^{\delta}}{\Delta}
\bar{\Omega} \varphi_*^2.
\end{eqnarray}

By using the obtained mass and particle number, the binding energy can be defined.
The value of BTZ mass $M_{BTZ}$ is negative when the matter of boson star is a little.
In particular, in the limit of ``no matter'', $M_{BTZ}$ approaches $-1/(8G)$.
Thus we will take the binding energy as
\begin{equation}
E=M_{BTZ}+\frac{1}{8G}-mN.
\end{equation}

\begin{figure}[p]
\centering
\epsfbox{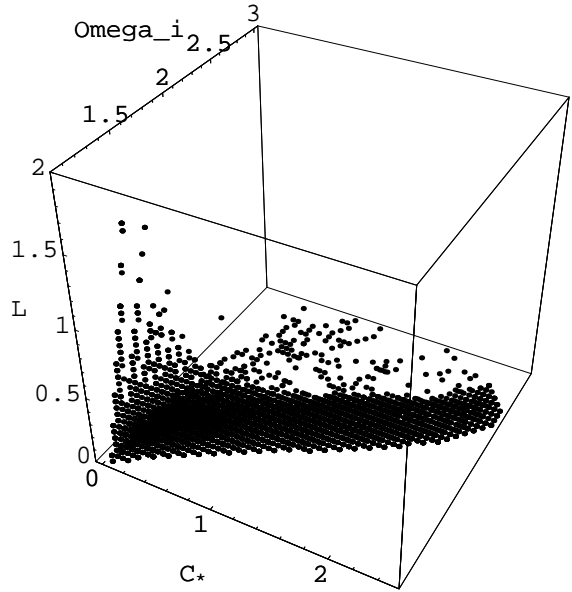}
\caption{The region of the stable solution.
These points are shown in the region of stable and possible solutions.}
\label{fig2}
\end{figure}

The binding energy must be negative for the stable boson star.
Therefore the region of stable solution is restricted.
The region of the stable and possible solution is shown in Fig.~\ref{fig2}.
In this parameter region, the stable boson star can exist.
In Fig.~\ref{fig2},
the maximum value of the cosmological constant is
$C_* = C_{*crit} \approx 2.5268$ for $L\to 0$.
For a nonrotating  boson star, the maximum value of
the cosmological constant is also
$C_* = C_{*crit} \approx 2.5268$ \cite{rf:3}.
This is trivial, because a rotating solution approaches
a nonrotating solution when $L\to 0$.
From Fig.~\ref{fig2}, one can find the region where the stable boson star can exist is restricted.

These parameters, particularly $L$ and $\bar{\Omega}_i$, is not the meaningful physical parameters.
Thus we rewrite these parameters in the meaningful physical parameters.
Instead of these, we will use the angular momentum of the boson star
$J_*$, the mass of the boson star $M_{BTZ}$
and the cosmological constant $C_*$
as the meaningful physical parameters.

We will consider the angular momentum of the boson star. 
Since the external solution of the boson star corresponds to BTZ solution,
$\Delta$ is given as
\begin{eqnarray}
\Delta&=&Cr^2-8GM_{BTZ}+\frac{(8GJ)^2}{4r^2} \nonumber \\
&=&C_*r_*^2-8GM_{BTZ}+\frac{(8GJ_*)^2}{4r_*^2},
\end{eqnarray}
where $J^2_*=(m^2/\Lambda)J^2$ is the angular momentum of the boson star.
Thus the angular momentum of the boson star $J_*$ is given by
\begin{equation}
8GJ_*=\frac{r^3_{*o}}{L}e^{\delta_o}
\left( \bar{\Omega}'_o \right).
\end{equation}

\begin{figure}[t]
\centering
\epsfbox{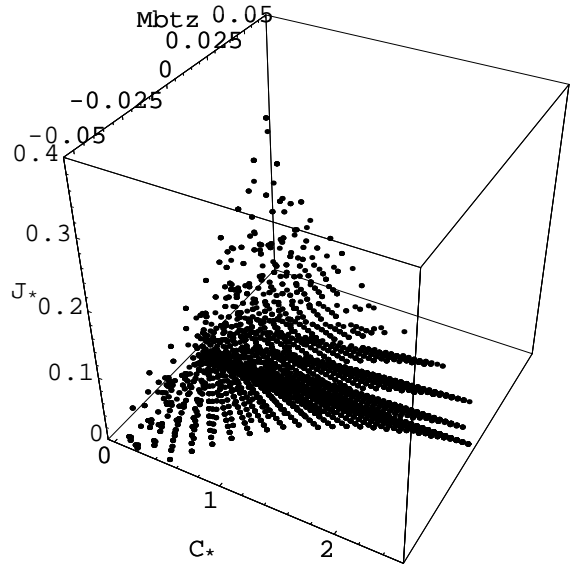}
\caption{The region of the stable boson star showing with
the angular momentum of boson star $J_*$,
the mass of boson star $M_{BTZ}$
and the cosmological constant $C_*$}
\label{fig3}
\end{figure}

\begin{figure}[htbp]
\centering
\epsfxsize=12cm
\epsfbox{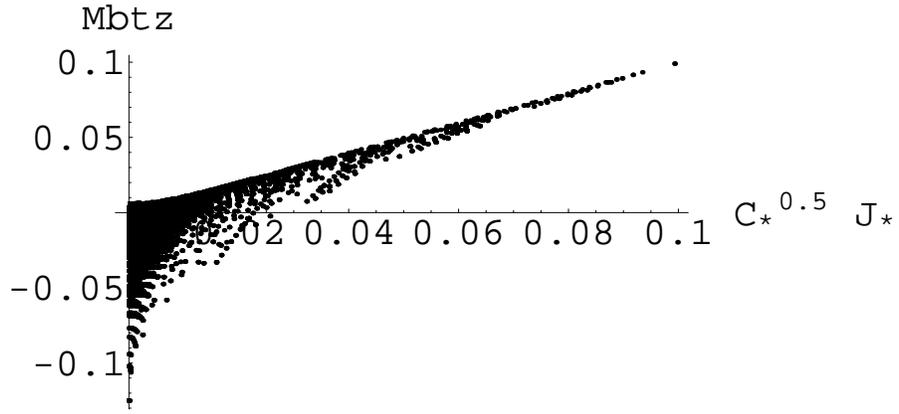}
\caption{The region of the stable boson star showing with $\sqrt{C_*} J_*$ and $M_{BTZ}$}
\label{fig4}
\end{figure}

Using these parameter, $J_*$, $M_{BTZ}$ and $C_*$,
the region of the stable boson star is shown in Fig.~\ref{fig3}.
Simplifying Fig.~\ref{fig3}, the region of the stable boson star
 is shown  with
$\sqrt{C_*} J_*$ and $M_{BTZ}$ in Fig.~\ref{fig4}.

In the region of Fig.~\ref{fig4}, the stable boson star can exist.
To understand Fig.~\ref{fig4}, we consider the case of the black hole physics.
In the case of the black hole, it is known
that the black hole with the larger angular momentum than its mass
has the naked singularity and cannot be a usual black hole.
In an opposite case, as the larger mass than the angular momentum, 
the black hole can be a usual black hole.
Thus, if the boson star become the black hole adiabatically, the boson star with 
$\sqrt{C_*} J_*<M_{BTZ}$ can become the usual black hole.

In Fig.~\ref{fig4}, there is the relation $\sqrt{C_*} J_*=M_{BTZ}$
when the mass is large.
This fact is interesting.
Because the black hole, which has such the same value of the angular
momentum and the mass, is the extreme black hole.
Thus, the boson star with the large mass can certainly become the extreme black hole.
In the case of boson stars, how the behavior of the extreme
black hole is appeared is very interesting.
But, if the mass of boson star is larger, the singularity is more often caused and
we cannot calculate the field equations numerically.
Therefore the numerical analysis of boson stars with the sufficient large mass is very difficult.

To summarize, we have obtained the numerical solutions describing the rotating boson stars
with a very large self-coupling constant in (2+1) dimensions.
We found that the rotating boson star has
a vacuum hole in the center of the star.
And the region of the solution for the stable boson star is shown.
There is the maximum value of the cosmological constant
$C_* = C_{*crit} \approx 2.5268$, where $C_*=(\lambda/(8\pi G m^4))C$.
This corresponds to the nonrotating case.
When the mass of boson star is large, there is the relation between the angular
momentum and the mass of boson star, $\sqrt{C_*} J_*=M_{BTZ}$.

The future plan for study of boson stars is the following.
The analysis of boson star with the actually finite value
of self-coupling is of much interest and will be necessary.
More general cases including such as a $|\varphi|^6$ coupling
may exhibit more complicated results, but the analysis of them
can be carried out in the same manner as in the present study.
On the other hand, the model of the spinning boson star
 in (3+1) dimensions has been studied~\cite{Ryan}.
The similarity and the difference, between
 our model and the model in the other dimensions, must be further studied.
Particularly, it is interesting to study the shape of the boson star with a large
angular momentum in various dimensions.

%%%%%%%%%%%%%%%%%%%%%%%%%%%%%%%%%%%%%%%%%%%%%%%%%%%%%%%%%%%%%%%%%%%%%%
\section*{Acknowledgement}
The authors would like to thank to S.~Hirenzaki for useful advice.
%%%%%%%%%%%%%%%%%%%%%%%%%%%%%%%%%%%%%%%%%%%%%%%%%%%%%%%%%%%%%%%%%%%%%%

%%%%%%%%%%%%%%%%%%%%%%%%%%%%%%%%%%%%%%%%%%%%%%%%%%%%%%%%%%%%%%%%%%%%%%
%%%References
%%%%%%%%%%%%%%%%%%%%%%%%%%%%%%%%%%%%%%%%%%%%%%%%%%%%%%%%%%%%%%%%%%%%%%

%%%%%%%%%%%%%%%%%%%%%%%%%%%%%%%%%%%%%%%%%%%%%%%%%%%%%%%%%%%%%%%%%%%%%%
%%%FIGURES
%%%%%%%%%%%%%%%%%%%%%%%%%%%%%%%%%%%%%%%%%%%%%%%%%%%%%%%%%%%%%%%%%%%%%%
%%%%%%%%%%%%%%%%%%%%%%%%%%%%%%%%%%%%%%%%%%%%%%%%%%%%%%
%%%%%%%%%%%%%%%%%%%%%%%%%%%%%%%%%%%%%%%%%%%%%%%%%%%%%%%

%%%%%%%%%%%%%%%%%%%%%%%%%%%%%%%%%%%%%%%%%%%%%%%%%%%%%%
%%%%%%%%%%%%%%%%%%%%%%%%%%%%%%%%%%%%%%%%%%%%%%%%%%%%%%%
\hspace{5cm}
%%%%%%%%%%%%%%%%%%%%%%%%%%%%%%%%%%%%%%%%%%%%%%%%%%%%%%
%%%%%%%%%%%%%%%%%%%%%%%%%%%%%%%%%%%%%%%%%%%%%%%%%%%%%%%

%%%%%%%%%%%%%%%%%%%%%%%%%%%%%%%%%%%%%%%%%%%%%%%%%%%%%%
%%%%%%%%%%%%%%%%%%%%%%%%%%%%%%%%%%%%%%%%%%%%%%%%%%%%%%%

%%%%%%%%%%%%%%%%%%%%%%%%%%%%%%%%%%%%%%%%%%%%%%%%%%%%%%%%%%%%%%%%%%%%%%
\end{document}